# Study the effects of metallic ions on the combination of DNA and histones with molecular combing technique

Yu-Ying Liu, Peng-Ye Wang[*], Shuo-Xing Dou, Ping Xie, Wei-Chi Wang, Hua-Wei Yin

*Laboratory of Soft Matter Physics, Institute of Physics, Chinese Academy of Sciences, Beijing 100080, China*

**Abstract** The effects of monovalent ($Na^+$, $K^+$) and divalent ($Mg^{2+}$, $Ca^{2+}$, $Mn^{2+}$) ions on the interaction between DNA and histone are studied using the molecular combing technique. λ-DNA molecules and DNA-histone complexes incubated with metal cations ($Na^+$, $K^+$, $Mg^{2+}$, $Ca^{2+}$, $Mn^{2+}$) are stretched on hydrophobic surfaces, and directly observed by fluorescence microscopy. The results indicate that when these cations are added into the DNA solution, the fluorescence intensities of the stained DNA are reduced differently. The monovalent cations ($Na^+$, $K^+$) inhibit binding of histone to DNA. The divalent cations ($Mg^{2+}$, $Ca^{2+}$, $Mn^{2+}$) enhance significantly the binding of histone to DNA and the binding of the DNA-histone complex to the hydrophobic surface. $Mn^{2+}$ also induces condensation and aggregation of the DNA-histone complex.

Recently, DNA stretching has attracted much attention. The study of DNA is being greatly advanced by the fluorescence microscopy technique at the single molecule level[1-2]. Several physical methods have been employed to stretch single DNA molecules, such as magnetic tweezer[3-5], laser tweezer[6-9], micropipette[10] and molecular combing method[11-13]. Up to now, we have studied the force-induced melting phenomenon of single DNA molecules with molecular combing method and fluorescence microscopy[14]. Some molecular combing results of the DNA-histone complex are reported[15]. Histone is a basic protein and plays an important role in DNA condensation. In a certain range it can fold DNA molecule into higher ordered structure. Eukaryotic genes do not exist naturally as naked DNA molecules. Instead, they combine with some proteins, especially the basic proteins called histones, to form a substance known as chromatin. Using Brownian dynamics simulation, we numerically studied the interaction of DNA with histones and proposed an octamer-rotation model to describe the process of nucleosome formation and nucleosome disruption under stretching[16,17]. Because both DNA and histone are biomacromolecules, the interaction between them is intricate. There are many factors to influence this interaction. For example, metal cations can alter DNA structure[18] or induce DNA condensation[19]. The effects of metal cations on DNA structure have been investigated by a variety of techniques, including UV-visible spectrophotometry, circular dichroism spectroscopy, NMR spectroscopy, and sedimentation equilibrium measurements[20]. In this paper, the effects of monovalent ($Na^+$, $K^+$) and divalent ($Mg^{2+}$, $Ca^{2+}$, $Mn^{2+}$) ions on the interaction between DNA and histone are studied using the molecular combing technique. DNA molecules and DNA-histone complexes are stretched and aligned on hydrophobic surfaces, and directly observed by fluorescence microscopy. The effect of metal cations on the interaction between DNA and histone can be analyzed by observing the length and the density of the stretched DNA-histone complex. This method is helpful to investigate the interaction between DNA and histone in single molecular level[21]. Up to date, to study the interaction between DNA and histone from a view of physics is a challenging subject, and therefore it is still an open problem awaiting theoretical and experimental approach.

---

[*] Corresponding author. E-mail: pywang@aphy.iphy.ac.cn

**1 Materials and methods**

(i) Experimental materials and equipments. Lambda DNA was purchased from Sino-American Biotechnology Company (China). Fluorescent dye, oxazole yellow dimmer (YOYO-1) was purchased from Molecular Probes Company (USA). Bis-Tris and histone were purchased from Sigma Company (USA). NaCl, KCl, $MgCl_2$, $CaCl_2$ and $MnCl_2$ were all purchased from Sigma Company (USA). All buffers were made with 18.2MΩ·cm water purified through the Milli-Q Water Purification System (Millipore Corporation, France). Fluorescently stained DNA molecules were observed by using an inverted fluorescence microscope (IX-70; Olympus). The images were captured by a cooled CCD camera (CoolSNAP-HQ, Roper Scientific). A MetaMorph software (Universal Imaging Corporation) was used for the system control, data acquisition and data processing.

(ii) Surface treatment. Quartz plates (21× 42 $mm^2$) were immersed in hot NaOH (0.5 mol/L) for about 10 min, and then rinsed thoroughly in high-purity water (Millipore S.A., France). Quartz surfaces were rendered hydrophobic by coating the surfaces with PMMA. A droplet (0.2 ml~0.3 ml) of PMMA (560F, Japan) in chloroform [10% (wt/wt)] was dripped down onto the center of a cleaned quartz surface, which was mounted horizontally on the spin-coating machine using a double-sided tape, and spread by spin-coating at 5500 revolutions per minute for 1 min. After spin-coating, the PMMA film was formed evenly on the whole surface. The quartz plate was then baked at 145°C for ~ 30 min and then stored at room temperature in a dust-free environment. During the preparation of PMMA film, it was very important to keep the uniformity of the PMMA solution and the cleanness of the surface.

(iii) DNA preparation. Lambda DNA (48.5 kb) was stained with a fluorescent dye, oxazole yellow dimmer (YOYO-1, Molecular Probes) at a ratio of ten base pairs per dye molecule (bp/dye = 10) by mixing DNA sample with a specific volume of freshly prepared 0.1 μmol/L dye solution (10 mmol·$L^{-1}$ Tris/1 mmol·$L^{-1}$ EDTA buffer, pH 8.0). The DNA/YOYO-1 solution was incubated for approximately 30 min in the dark, and then was diluted to 6.5pmol/L in a 50mmol/L Bis-Tris buffer (pH 6.6, Sigma).

(iv) DNA-metal cation complexes preparation. $DNA-Na^+$ preparation procedure was that DNA/YOYO-1 solution（6.5pmol/L）was incubated with NaCl at 37℃ for about 30 min. The concentrations of NaCl were 30, 50, 100, 150mmol/L respectively. $DNA-K^+$ preparation procedure was the same as that of $DNA-Na^+$ (NaCl was substituted by KCl). $DNA-Mg^{2+}$ preparation procedure was that DNA/YOYO-1 solution (6.5pmol/L) was incubated with $MgCl_2$ at 37℃ for about 30 min. The concentrations of $MgCl_2$ were 1, 2, 3, 5mmol/L respectively. The preparation procedure of $DNA-Ca^{2+}$ and $DNA-Mn^{2+}$ were the same as that of $DNA-Mg^{2+}$. The concentrations of $MnCl_2$ and $CaCl_2$ were 1, 2, 3, 5mmol/L respectively.

(v) DNA-metal cation-histone complexes preparation.

(1) $DNA-Na^+$-histone complexes solution. Histone was diluted by Bis-Tris (50mmol/L,pH~ 6.6) buffer. DNA (6.5pmol/L) and diluted histone were incubated (the molar ratio of DNA to

histone was 1:100) together at 37℃ for 30 min and this was as the comparison group. To prepare DNA-Na$^+$-histone complexes solution, DNA, NaCl and histone were incubated together at 37℃ for 30 min. The concentrations of NaCl were 30, 50, 100, 150mmol/L respectively. Then a droplet (1~ 2μl) of the reaction solution was deposited onto a PMMA surface, with the drying of the droplet, DNA molecules originally bound to the surface with one extremity were extended and immobilized on the dried surface, and these combed DNA were observed with fluorescence microscopy. The preparation procedure of DNA-K$^+$-histone complexes solution was the same as that of above.

(2) DNA-Mg$^{2+}$-histone complexes solution. Under the identical experiment conditions, DNA, MgCl$_2$ and histone were incubated together at 37℃ for 30 min. The DNA-histone complexes group was as comparison group (the molar ratio of DNA to histone was 1:100). The concentrations of MgCl$_2$ were 1, 2, 3, 5mmol/L respectively. The experiment procedure of DNA-Ca$^{2+}$-histone (the molar ratio of DNA to histone was 1:50) and DNA-Mn$^{2+}$-histone complexes solution (the molar ratio of DNA to histone was 1:300) were the same as that of DNA-Mg$^{2+}$-histone complexes. The concentrations of MnCl$_2$ and CaCl$_2$ were 1, 2, 3, 5mmol/L respectively.

(vi) DNA imaging by fluorescence microscopy. YOYO-1 has an excitation maximum at 491 nm and emission maximum at 509 nm, that is, YOYO-1 molecules emit green fluorescence under the excitation of blue light, then DNA molecules can be seen by observing the fluorescence of YOYO-1. A 100W mercury lamp was used in combination with a U-MWB excitation cube (BP450-480, DM500, BA515). Fluorescently stained DNA molecules were observed by using an inverted optical microscope (IX-70; Olympus) by epifluorescence with a 20μ objective. The images were captured by a cooled CCD camera (CoolSNAP-HQ; Roper Scientific, Inc.). The CCD acquisition time we used was three seconds.

**2 Results**

(i) Mg$^{2+}$, Ca$^{2+}$, Mn$^{2+}$ enhanced DNA binding to PMMA surfaces significantly. We had observed that Mg$^{2+}$ improved DNA binding efficiency significantly in the absence of histone in previous experiment [14]. With the increasing of Mg$^{2+}$, the density of combed DNA increased. The effect of Ca$^{2+}$ and Mn$^{2+}$ on combed DNA was also investigated, together with that of no metal ions for comparison. The results showed that when adding CaCl$_2$ or MnCl$_2$ (1, 2, 3, 5mmol/L) in DNA solution the binding efficiency was significantly improved in contrast to comparison group (see Figures 1, 2). This suggested that Mg$^{2+}$, Ca$^{2+}$, and Mn$^{2+}$ might have similar character in binding DNA. We found that these three metal ions enhanced DNA binding to PMMA surface. This should be due to the fact that these ions increase the hydrophobic interaction between DNA and the PMMA surface by increasing the molal surface tension of water.

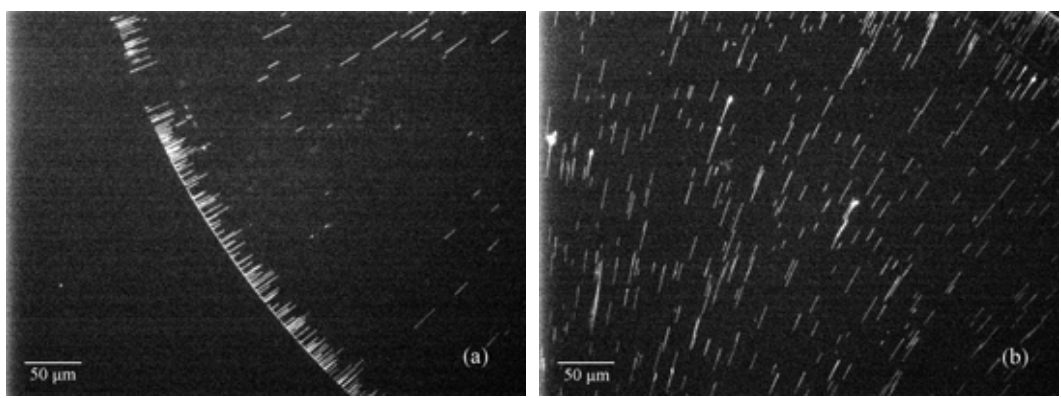

Fig. 1. CCD images of single combed DNA molecules with buffers of 6.5pmol/L DNA at pH 6.6, and different concentrations of added $Ca^{2+}$: (a) 0mmol/L (comparison); (b) 3mmol/L. [Bars: 50 μm (a, b).]

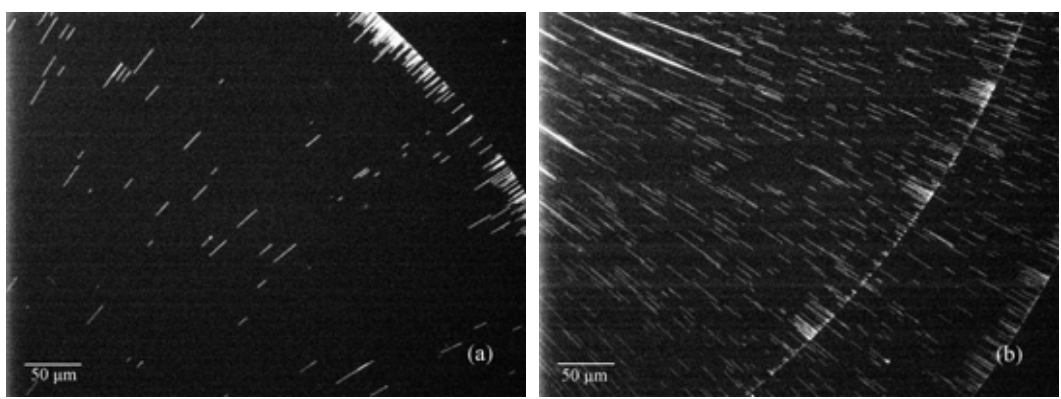

Fig. 2. CCD images of single combed DNA molecules with buffers of 6.5pmol/L DNA at pH 6.6, and different concentrations of added $Mn^{2+}$: (a) 0mmol/L (comparison); (b) 3mmol/L. [Bars: 50 μm (a, b).]

(ii) $Na^+$, $K^+$, $Mg^{2+}$, $Ca^{2+}$ and $Mn^{2+}$ reduced the fluorescence intensity of DNA. Our experimental results showed that these metal ions induced DNA fluorescence intensity attenuation differently. $Na^+$ and $K^+$ reduced DNA fluorescence intensity weakly. When the concentration of $Na^+$ or $K^+$ was 30mmol/L, DNA-$Na^+$ or DNA-$K^+$ fluorescence intensity had no difference with that of comparison; only when the concentration of $Na^+$ or $K^+$ was higher than 100mmol/L, the fluorescence intensity of DNA-$Na^+$ or DNA-$K^+$ was reduced obviously. The effect of $Mg^{2+}$ and $Ca^{2+}$ on the attenuation of DNA fluorescence intensity was much stronger than that of monovalent cations at a given concentration. When [$Mg^{2+}$] was 5mmol/L, the fluorescence intensity of combed DNA molecules was very dim, so the maximum concentration of divalent ions was no more than 5mmol/L. Particularly, $Mn^{2+}$ had the strongest effect on the attenuation of DNA fluorescence intensity among these metal ions. Even [$Mn^{2+}$] was as low as 1mmol/L, the fluorescence intensity of DNA was too weak to be observed clearly.

We measured the mean fluorescence intensity of DNA and DNA-ion complexes by MetaMorph software. The results showed that when [$Na^+$] was 150mmol/L, the mean fluorescence intensity of DNA-$Na^+$ was 0.79 times of that of its comparison. Similarly, for 150mmol/L [$K^+$], DNA-$K^+$ mean fluorescence intensity was 0.84 times of that of its comparison. In the case of 3mmol/L [$Mg^{2+}$], [$Ca^{2+}$] and [$Mn^{2+}$], the mean fluorescence intensities of DNA-$Mg^{2+}$, DNA-$Ca^{2+}$ and DNA-$Mn^{2+}$ were 0.71, 0.63 and 0.45 times of that of their comparisons respectively (see Figure 3). On the other hand, under the condition of higher ionic strength for the same ion, the fluorescence intensity of the stained DNA was much dimmer (data not shown).

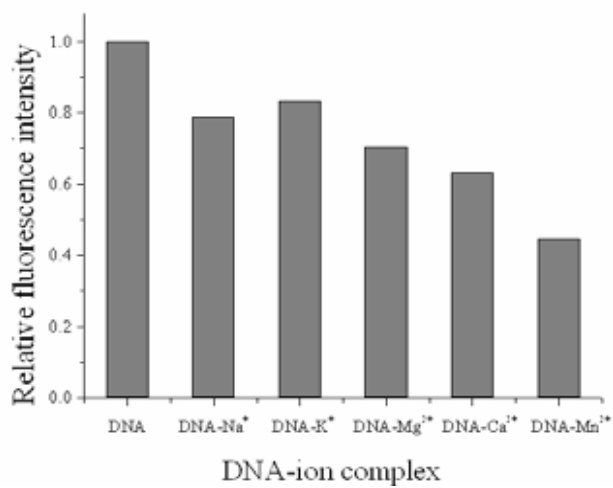

Fig. 3. Relative fluorescence intensity with buffers of 6.5pmol/L DNA and different concentrations of added ions: [$Na^+$], [$K^+$] =150mmol/L; [$Mg^{2+}$], [$Ca^{2+}$], [$Mn^{2+}$] =3mmol/L

From above, we deduced that there were two binding modes of YOYO-1 on DNA: intercalation and external binding (groove association). The intercalation was the predominant binding mode. It was proposed that monovalent cations located in the grooves of B-DNA[22]. So the positively charged Na and K ions might compete against YOYO-1 for groove sites, thus reducing the YOYO-1 binding efficiency. Still there were certain numbers of YOYO-1 molecules intercalating into base pairs of DNA, so DNA molecules with bright fluorescence could also be seen.

Raman spectra revealed that transition metal cations ($Mn^{2+}$, $Co^{2+}$, $Cu^{2+}$) induced great structure changes in B-DNA and the change was stronger for longer DNA. The interactions between these transition metal cations and the DNA bases were stronger than interactions between them and the DNA phosphates [20]. In case of $Mn^{2+}$ the interaction with bases was so strong that it disturbed YOYO-1 intercalating into bases. In fact, the fluorescence intensity was much dimmer in the presence of $Mn^{2+}$. This phenomenon proved that $Mn^{2+}$ interacted with bases dominantly. Some researches indicated that alkaline earth metals ($Mg^{2+}$, $Ca^{2+}$) had higher affinities for the phosphates, and the decreasing order for the strength of metal-ion interaction with DNA bases was $Mn^{2+}$ > $Ca^{2+}$ > $Mg^{2+}$ [20]. In our experiment, we found that the effect of $Mg^{2+}$ and $Ca^{2+}$ on the fluorescence attenuation of DNA was stronger than that of monovalent cations, but weaker than that of $Mn^{2+}$. This indicated that the interaction of $Mn^{2+}$ with bases was much stronger than that of $Ca^{2+}$ and $Mg^{2+}$. In summary, the presence of the ions caused reduction of the binding efficiency of YOYO-1 with DNA, the decreasing order for the strength of effect was $Mn^{2+}$>$Ca^{2+}$~ $Mg^{2+}$> $K^+$ ~ $Na^+$ (see Figure 3).

In addition, it was claimed[23] that the quantum yield also dropped in the order $Cu^{2+}$> $Mn^{2+}$> $Mg^{2+}$

when divalent cations were added. This implicated that the fluorescence change not only originated from the decreased number of the bound fluorescence ligands, but also related to the energy transfer efficiency. The stronger the metal ions interacted with the bases, the lower the energy transfer was[23].

From above we concluded that the binding modes of monovalent and divalent cations on DNA had obvious diferences. Especially for divalent cations ($Mn^{2+}$, $Ca^{2+}$, $Mg^{2+}$), they probably shared some similar binding mechanism which enhanced DNA binding capacity to PMMA surface. But $Mn^{2+}$, $Ca^{2+}$ and $Mg^{2+}$ ions had different binding sites on DNA, which induced fluorescence attenuation differently. Using molecular combing and fluorescence microscopy methods, we observed large number of combed DNAs directly and validated metal ions binding properties to some extent.

(iii) $Na^+$ and $K^+$ inhibited the binding of histone to DNA. Without ions, the combed DNA-histone complexes (comparison group, see Figure 4b) were denser than that of single DNA molecules on PMMA surface (see Figure 4a). Our previous experimental results[16] showed that a large number of histones combined to DNA. Moreover, with the increasing of histone, the combed DNA-histone complexes became denser and their lengths became shorter. In certain cases the DNA-histone complexes condensed and could not be stretched fully. However, when DNA, histone and NaCl were incubated together for 30 min, the stretched DNA-histone-$Na^+$ complexes (see Figure 4c) became sparser than that of comparison (see Figure 4b), and for different concentrations of $Na^+$ this phenomenon was similar. This result showed that $Na^+$ occupied some binding sites on DNA for histone, so that fewer histones combined to DNA, thus inhibiting DNA and histone binding.

Under the same experiment condition as for $Na^+$, the effect of $K^+$ on the interaction of DNA and histone was studied. $K^+$ and $Na^+$ had similar effect according to the distribution of the combed DNA-histone-$K^+$ (see Figure 4d).

(iv) $Mg^{2+}$, $Ca^{2+}$ and $Mn^{2+}$ enhanced the binding of histone to DNA significantly. At the given concentration ratio of histone to DNA, we got the combed DNA-histone complexes (comparison group, see Figure 5b) which were denser than that of single combed DNA molecules (see Figure 5a), and then we studied the effect of $MgCl_2$ on the interaction of histone and DNA. The results showed that after DNA, histone and $MgCl_2$ were incubated together for 30 min, the stretched DNA-histone-$Mg^{2+}$ complexes became denser together with obviously shorter length than that of comparison, even some DNA-histone-$Mg^{2+}$ complexes appeared as small bright dots on PMMA surface (see Figure 5 c, d). This suggested that the binding of DNA and histone became more tightly when adding $Mg^{2+}$ into reaction buffer, so complexes molecules could not be fully stretched. In addition, the binding capacity of DNA-histone to PMMA surface was enhanced with adding $MgCl_2$, and this effect was observed for different concentrations of $Mg^{2+}$ (1, 2, 3, 5mmol/L). We also found that DNA condensation was more obvious with higher concentration of $MgCl_2$.

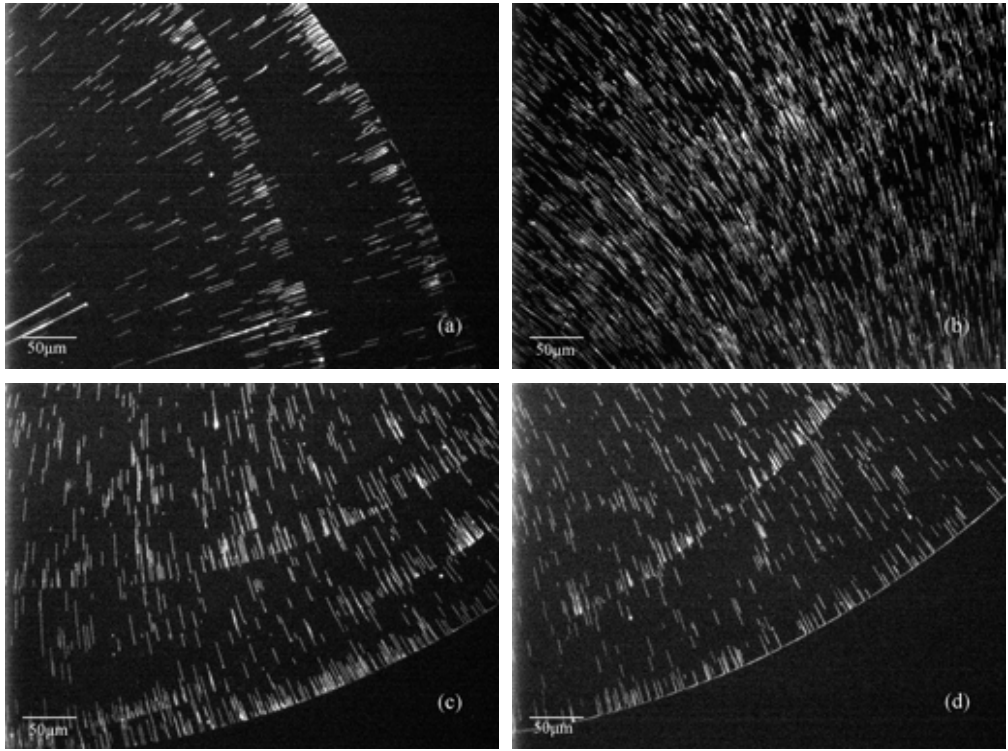

Fig.4. CCD images of single combed DNA, DNA-histone and DNA-histone-ion complexes: [DNA] = 6.5pmol/L. (a) [Histone] = 0,(b) [Histone] = 650pmol/L (comparison group), (c) [Histone] = 650pmol/L, [NaCl] = 30mmol/L,(d) [Histone] = 650pmol/L, [KCl] = 50mmol/L. [Bars: 50μm (a, b, c and d)]

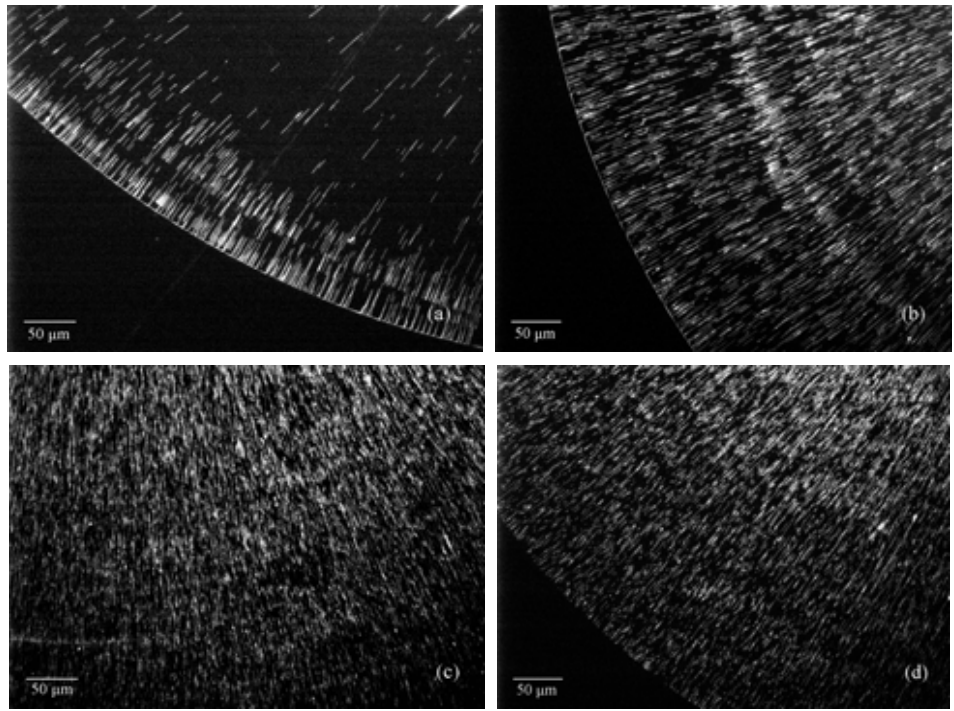

Fig.5. CCD images of single combed DNA, DNA-histone and DNA-histone-$Mg^{2+}$ complexes: [DNA] = 6.5pmol/L. (a) [Histone] = 0, (b) [Histone] = 650pmol/L (comparison group), (c) [Histone] = 650pmol/L, [$MgCl_2$] = 1mmol/L, (d) [Histone] = 650pmol/L, [$MgCl_2$] = 2mmol/L.

CaCl$_2$ could also enhance histone binding to DNA and strengthen the binding capacity of DNA-histone complexes to PMMA surface, and DNA-histone complexes condensation was observed (see Figure 6).

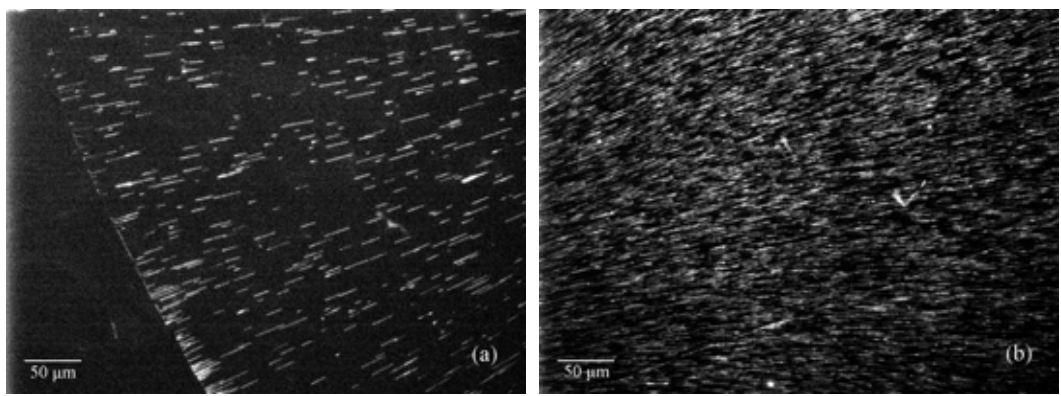

Fig.6. CCD images of DNA-histone and DNA-histone-Ca$^{2+}$ complexes: [DNA] = 6.5pmol/L, [Histone] = 325pmol/L. (a) [CaCl$_2$] = 0(comparison group), (b) [CaCl$_2$] = 3mmol/L.

MnCl$_2$ made DNA-histone complexes more difficult to be stretched. This indicate that DNA and histone bind each other more tightly in the presence of Mn$^{2+}$. In addition, when adding Mn$^{2+}$, DNA-histone complexes appeared as many bright dots distributed on the PMMA surface (see Figure 7), and this also showed DNA-histone complexes condensed tightly and could not be stretched. From above we concluded that these three divalent metal cations could enhance the interaction of DNA and histone, particularly, the effect of Mn$^{2+}$ was the most significant.

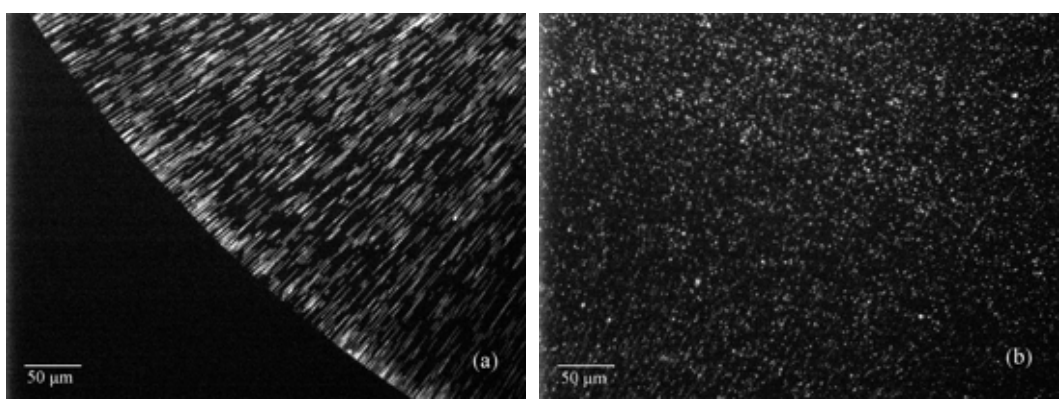

Fig.7. CCD images of DNA-histone and DNA-histone-Mn$^{2+}$ complexes: [DNA] = 6.5pmol/L, [Histone] = 1.95nmol/L. (a) [MnCl$_2$] = 0 (comparison group), (b) [MnCl$_2$] = 0.5mmol/L.

It was observed by atomic force microscopy that the divalent cations (Mg$^{2+}$, Mn$^{2+}$, Cu$^{2+}$) could induce DNA condensation [23]. Our results using molecular combing method validated that histone could also induce DNA condensation. When divalent cation, histone and DNA incubated together for some time, the combed complexes density increased significantly and each single complex condensed more seriously. Comparing to the case of DNA incubated with divalent cations or histone, there were more histones and divalent cations binding to DNA. That is, when divalent cations and histone occurred at the same time they could cooperate with each other. It might be that in the course of divalent cations binding to DNA, divalent metal cations altered DNA structure which favored the interaction of DNA and histone.

The effect of metal ions ($Mn^{2+}$, $Ca^{2+}$, $Cu^{2+}$) on the DNA structure was previously studied at different relative humidities (5–98%) by IR-spectroscopy. The results suggested the interaction of the ions both with the DNA phosphate groups and with the nucleic bases. The formation of the secondary structure of DNA complexes with metal ions was shown to take place at a greater number of water molecules bound to the polymer than it was the case of DNA without ions[24]. Once a condensed nucleus was formed, the hydration force would drive the remaining part to wind around it [23]. These mechanisms might be the reason for divalent metal cations accelerating the interaction of histone and DNA. But this binding mechanism was related to the kind of cations and the binding sites of metal ions. These three divalent cations have different binding sites. For instance, $Mg^{2+}$ and $Ca^{2+}$, typical divalent cations mainly binding to phosphate, could neutralize the negative charge of sugar-phosphate backbone; but $Mn^{2+}$ could coordinate with the electron-rich sites on the bases, thus these three divalent ions had different effect on the interaction of DNA and histone.

The effects of divalent cations on the DNA and chromatin conformation had been investigated by electric birefringence[25]. These observations were interpreted in terms of a specific organization of DNA in a compact, rigid structure, in the presence of $Mn^{2+}$ and $Cu^{2+}$, and a non-specific coiling in the presence of $Mg^{2+}$. Drastic conformational changes encountered by chromatin in the presence of $Mg^{2+}$ and $Mn^{2+}$ cations had also been evidenced through electric birefringence measurements. They were interpreted by the formation of a superhelical compact arrangement of nucleosome strings [25]. Our experiment that with adding divalent cations in DNA or DNA-histone solution, combed DNAs or DNA-histones were directly observed by molecular combing and fluorescence microscopy, validated some DNA binding properties of metallic ions to a certain extent.

Some previous research works pointed out that the metal cations exhibiting higher affinities for DNA bases would favor greater aggregation [20]. $Mn^{2+}$ could produce condensation of supercoiled DNA and provoke DNA condensation which may not only through an electrostatic mechanism [26]. In vitro experiment, because multivalent cations largely neutralize the high negative charge density of DNA, thus reducing interhelix electrostatic repulsion to make DNA condensation, and in Reference 26, Wilson and Bloomfield calculated that DNA condensation occurred when 89-90% of the DNA charge was neutralized. Condensing ligands might also act by crossbridging neighboring helices, perturbing hydration structure, or perturbing DNA helix structure [26]. In our experiment, we indeed observed that $Mn^{2+}$ readily induced DNA-histone complexes to aggregate together (see Figure 7b), and induced these complexes to bind to the hydrophobic surface tightly. The mechanism of DNA condensation and aggregation induced by $Mn^{2+}$ is still an open question and needs to be further studied.

**3 Summary**

We studied the effect of monovalent and divalent metallic cations on the interaction between DNA and histone by molecular combing method. The conclusion was as follows:
(1) Monovalent metal cations and histone competed to bind DNA when they existed together in DNA solution, thus monovalent metal cations inhibiting the combination of DNA and histone.

(2) When divalent metal cations and histone existed together in the solution, divalent metal cations enhanced histone binding to DNA and DNA binding to PMMA surfaces. And the combed DNA-histone complexes aggregated on the hydrophobic surface. It might be that divalent metal cations alter DNA structures which favor the interaction between DNA and histone.

(3) The binding mode of monovalent and divalent metal cations to DNA might be different.

Comparing to biochemical experiment, molecular combing method has the following virtues: simple procedure, evident experimental phenomenon, and few quantity sample etc. From this work, we conclude that the interaction of DNA and histone, with the effect of ions or other conditions of solution could be investigated qualitatively or quasi-quantitatively by molecular combing method.

**Acknowledgements** This work was supported by the National Natural Science Foundation of China (Grant Nos. 60025516 and 10334100)